\documentclass[a4paper]{PoS}
\usepackage{xspace} % For space after newcommand
\usepackage{aas_macros} % abbreviations des journaux
\usepackage[load-configurations = abbreviations,load=accepted]{siunitx} % unites
\DeclareSIUnit\ergon{erg}
\usepackage[numbers]{natbib}
\usepackage{subfig,graphicx}

\title{Looking for infrared counterparts of Fermi/LAT blazar candidates}

\ShortTitle{Looking for infrared counterparts of Fermi/LAT blazar candidates}

\author{\speaker{J. Lefaucheur}$^{\,a}$, C. Boisson$^a$, P. Goldoni$^b$, S. Pita$^b$\\
  \llap{$^a$}LUTH, Observatoire de Paris, PSL Research University, CNRS, Universit\'e Paris Diderot\\
  5 Place Jules Janssen, 92190 Meudon, France\\
  \llap{$^b$}APC, AstroParticule et Cosmologie, Universit\'{e} Paris Diderot, CNRS/IN2P3, CEA/Irfu,
  Observatoire de Paris, Sorbonne Paris Cit\'{e}\\
  10, rue Alice Domon et L\'{e}onie Duquet, 75205 Paris Cedex 13, France\\
  E-mail:  \email{julien.lefaucheur@obspm.fr}
}

\abstract{The Fermi/LAT telescope is an efficient blazar-detector in the MeV/GeV range. More than 1100 (900) blazars detected above 100 MeV (10 GeV) are clearly associated to BL Lacertae or Flat Spectrum Radio Quasar objects in the Fermi/LAT 3FGL catalogue. This number could significantly increase if multi-wavelength counterparts could be identified for the 573 3FGL blazars with unknown type, or even for the 1010 3FGL unassociated sources which are thought to be dominated by blazars, at least at high galactic latitude. Unfortunately, the size of the Fermi/LAT error box makes multi-wavelength follow-ups difficult.

We propose a method to associate ``blazar-like'' infrared counterparts, having coordinates with a precision of a few arcseconds, to  Fermi/LAT blazars and unassociated sources. To reach this goal, we built machine-learning classifiers based on the statistical differences of magnitude measurements obtained by the WISE satellite, between a sample of well-identified infrared blazars and samples of other types of infrared sources located in regions of the sky where no known blazar is present. We provide a list of potential infrared counterparts for 3FGL blazar candidates, along with the associated number of expected false positives. This study contributes to increase the number of well-identified extragalactic blazars and also provides promising blazar targets for the Cherenkov Telescope Array.}

\FullConference{35th International Cosmic Ray Conference -ICRC217-\\
		10-20 July, 2017\\
		Bexco, Busan, Korea}

\begin{document}

\section{Introduction}
\label{sec:intro}
% Interet des blazars
Blazars dominate the extragalactic sky above 100 MeV.
They are radio-loud active galactic nuclei (AGN) whose jet is quasi-aligned to the line of sight.
An important Doppler effect blue-shifts their spectra and increases their observed luminosity.
Their spectral energy distribution is characterised by a first bump peaking between the infrared
and the X-ray domain which is associated to synchrotron emission of relativistic electrons.
The high energy bump, in the MeV/TeV energy range, is usually associated to inverse-Compton
radiation in a leptonic scenario but might also be explained with hadronic scenarios.
In the very high energy range ($E \SI{\geq 100}{GeV}$), the current understanding of their
population and the improvement of the diffuse extragalactic background light measurement are nowadays
limited by the small number of detected blazars.

% Fermi/LAT et source unassoc
Since 2008, the Fermi/LAT telescope maps the sky above \SI{100}{\MeV}
with unmatched sensitivity and angular resolution in this frequency domain.
The LAT collaboration reported the detection of 3034 $\gamma$-ray sources \citep{Acero:2015aa}
among which 1717 are blazars, including 660 BL Lacertae (BL Lacs), 484
flat spectrum radio quasars (FSRQs) and 573 blazars of undetermined type (BCUs).
In addition, 1010 sources are still of unknown nature because of the lack
of firmly identified counterparts at other wavelengths.
The identification of these sources is not an easy task considering
the multiple candidate associations due to the large error localisation of the Fermi/LAT
and the incompleteness of counterpart catalogues.
It is expected that a significant fraction of the unassociated sources
are blazars since they are the dominant class of sources detected by the LAT.
Several studies \citep{Lefaucheur:2017aa, Saz-Parkinson:2016aa}
looked for blazar candidates among the unassociated sources with the help of
machine-learning classification methods and separation power
between different classes of sources extracted from Fermi/LAT catalogues.

% recherche contreparties IR, Massaro
The determination of possible counterparts to
the unassociated sources or the BCUs
will help to reveal their nature and simplify their
identification at other wavelengths.
Massaro et al. \cite{Massaro:2011aa} used the assumption that blazars occupy a
special position in the colour-colour diagram constructed with the
magnitudes measured with the WISE satellite \citep{Wright:2010aa}.
By building ``blazar regions'' with a selected sample of infrared blazars,
and by comparing the distance of the unassociated sources in the colour
space to these regions, one can identify potential candidates for
blazar-like counterparts \citep{Massaro:2013ab}.
However, this method is based only on a selected sample of known blazars
and does not consider the behaviour of other infrared source classes.
Therefore, it makes it hard to estimate the number of false positives.

% Plan
In this contribution, we propose a method to associate infrared counterparts
to high galactic latitude ($b\geq\SI{10}{\degree}$) $\gamma$-ray blazar candidates, along with
the number of expected false positives for each association.
We used three colors obtained from the four magnitudes extracted from
the WISE catalogue and a newly defined parameter miming the blazar
efficiency to produce infrared photons by synchrotron emission
to discriminate between ``blazar-like'' and non-blazar infrared counterparts.
Afterwards, we built a classifier with a sample of well-identified
infrared counterparts of blazars against a sample of infrared sources selected
in regions of the sky where no $\gamma$-ray blazar is present.
In order to estimate the number of false associations, we defined
different classes of associations according to the number of expected
false positives.

In Section~\ref{sec:samples} we introduce the different data samples
and the discriminant parameters.
Section~\ref{sec:train} presents the classifier construction and the procedure
to determine the infrared associations.
Results are shown in Section~\ref{sec:appli} and we conclude with a short
discussion in Section~\ref{sec:end}.

\section{Data samples and discriminant parameters}
\label{sec:samples}
% creer une classification a partir de differents sample
%% Different samples of infrared sources provided by the AllWISE Source
%% Catalogue\footnote{\protect\href{http://wise2.ipac.caltech.edu/docs/release/allwise/}{http://wise2.ipac.caltech.edu/docs/release/allwise/}}
%% were used to search for counterparts of high galactic latitude ($b>\SI{10}{\deg}$)
%% blazar candidates.
We used different samples of infrared sources, provided by the AllWISE Source
Catalogue\footnote{\protect\href{http://wise2.ipac.caltech.edu/docs/release/allwise/}{http://wise2.ipac.caltech.edu/docs/release/allwise/}}, to build a binary classifier to search for counterparts of high galactic latitude ($b>\SI{10}{\deg}$)
blazar candidates.
% dérougissement
We corrected the magnitudes of all sources for infrared Galactic extinction in the two
shorter wavelength filters $W_1$ ($\SI{3.4}{\micro \meter}$) and $W_2$ ($\SI{4.6}{\micro \meter}$)
using the Schlegel et al. \cite{Schlegel:1998aa}
data and the Indebetouw et al. \cite{Indebetouw:2005aa} extinction law.
In order to use reliable magnitude measurements, we only kept infrared
sources with:
\begin{itemize}
\item a signal/noise ratio greater or equal to 2 in all filters
\item a contamination and confusion flag only equal to ``0'', ``o'' or ``h'' in all filters\footnote{See \protect\href{http://wise2.ipac.caltech.edu/docs/release/allwise/expsup/sec2\_1a.html}{http://wise2.ipac.caltech.edu/docs/release/allwise/expsup/sec2\_1a.html} for further details.}
\item an ``extended'' flag less or equal to 1 corresponding to point-like source
\end{itemize}

\begin{figure}[t]
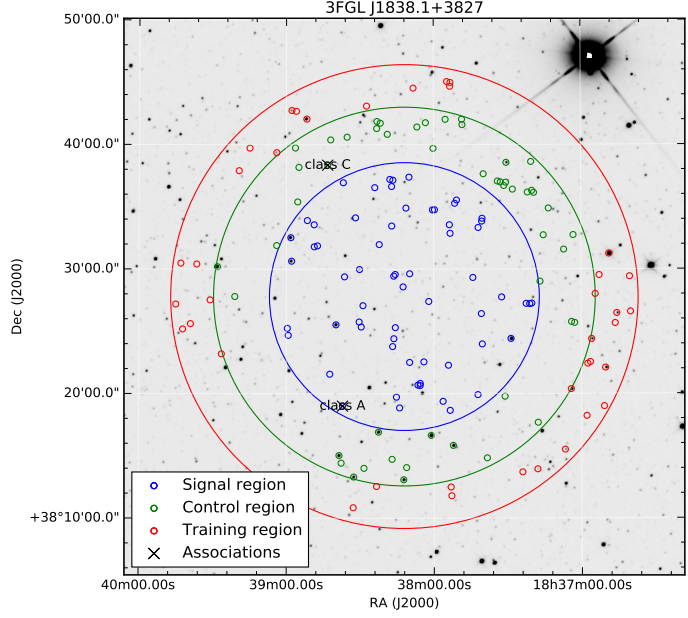

  \center 
  \includegraphics[width=.6\textwidth]{{{./3FGLJ1838.1+3827}}}
  \caption{Illustration of the different regions used in this study to select
    the infrared counterparts of a Fermi/LAT unassociated source.
    The image is a mosaic of individual frames from the WISE $W_1$ filter
    around the source.
  }
  \label{fig:map_wise}
\end{figure}
% Blazar signal
To create a sample of well-identified infrared blazars, we selected all the sources
from the 3LAC catalogue \citep{Ackermann:2015aa} labelled as BL Lac or FSRQ
belonging to the 3LAC clean sample (1018).
We further selected their infrared counterparts (754) in a circular region of radius
\SI{1}{\arcsecond} centered on the source position given by the 3LAC catalogue.
% fond IR (photo)
For the sample of non-blazar infrared sources, composed mainly of
stars, normal galaxies and QSOs, we selected and stacked all the sources in every annular
regions surrounding the 531 high-latitude 
unassociated sources (see Figure~\ref{fig:map_wise}), except the so-called
c-sources\footnote{The c-sources are considered to be potentially confused with galactic diffuse
  emission.}, of the 3FGL catalogue.
The inner and outer radius of the annular regions were respectively
set to $2 \sqrt{\theta_{95}}$ and $3 \sqrt{\theta_{95}}$, where
$\theta_{95}$ is the $\SI{95}{\percent}$ confidence level on the localisation of a Fermi/LAT source.
The stack, composed of $\sim$18900 sources, will be used as an estimate for the infrared
background sources.
%This stack is used as an estimate of the local infrared background 
%local estimates of the infrared background (see Figure~\ref{fig:map_wise}
%for an example).
% target
The target samples were defined as the high-latitude unassociated sources
and the high-latitude BCUs, without the c-sources.
For each of the sources in these two samples,
we selected all the infrared counterparts
in the circular region centered on the
LAT position , called the ``source region'', of radius $\theta_{95}$.
Furthermore, we selected the sources in an annular region surrounding each of the sources in
the two target samples, called the ``control regions'' and of the same area as the source region,
to search for potential blazar-like counterparts which are further away than the Fermi/LAT error boxes.
The inner and outer radius of the control regions were respectively
set to $\sqrt{\theta_{95}}$ and $2 \sqrt{\theta_{95}}$.

\begin{figure}[t]
  \centering
  \subfloat[]{
    \label{fig:c1c2}
    \includegraphics[width=0.33\columnwidth]{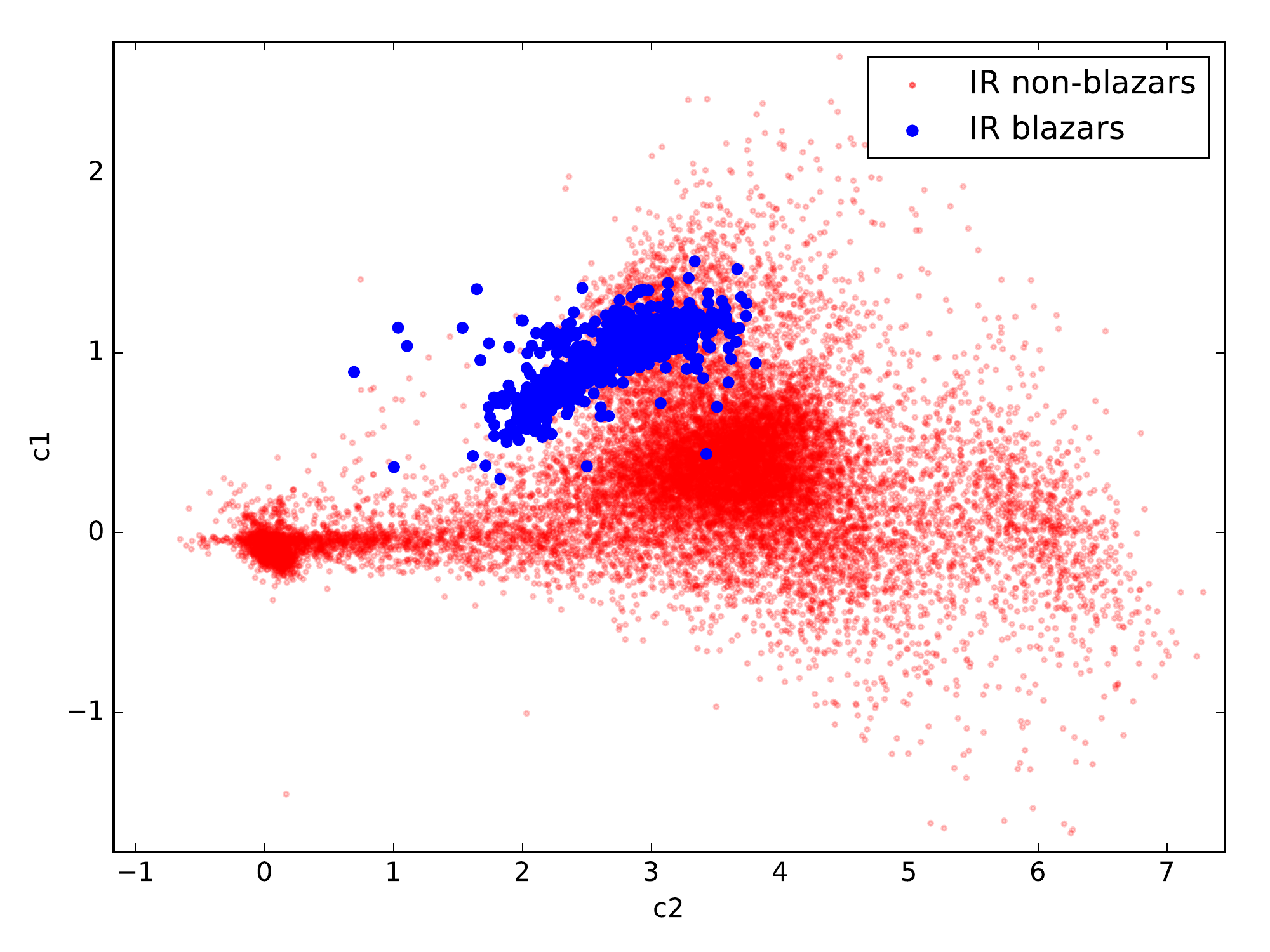}
  }
  %\hspace{0.1\linewidth}
  \subfloat[]{
    \label{fig:c2c3}
    \includegraphics[width=0.33\columnwidth]{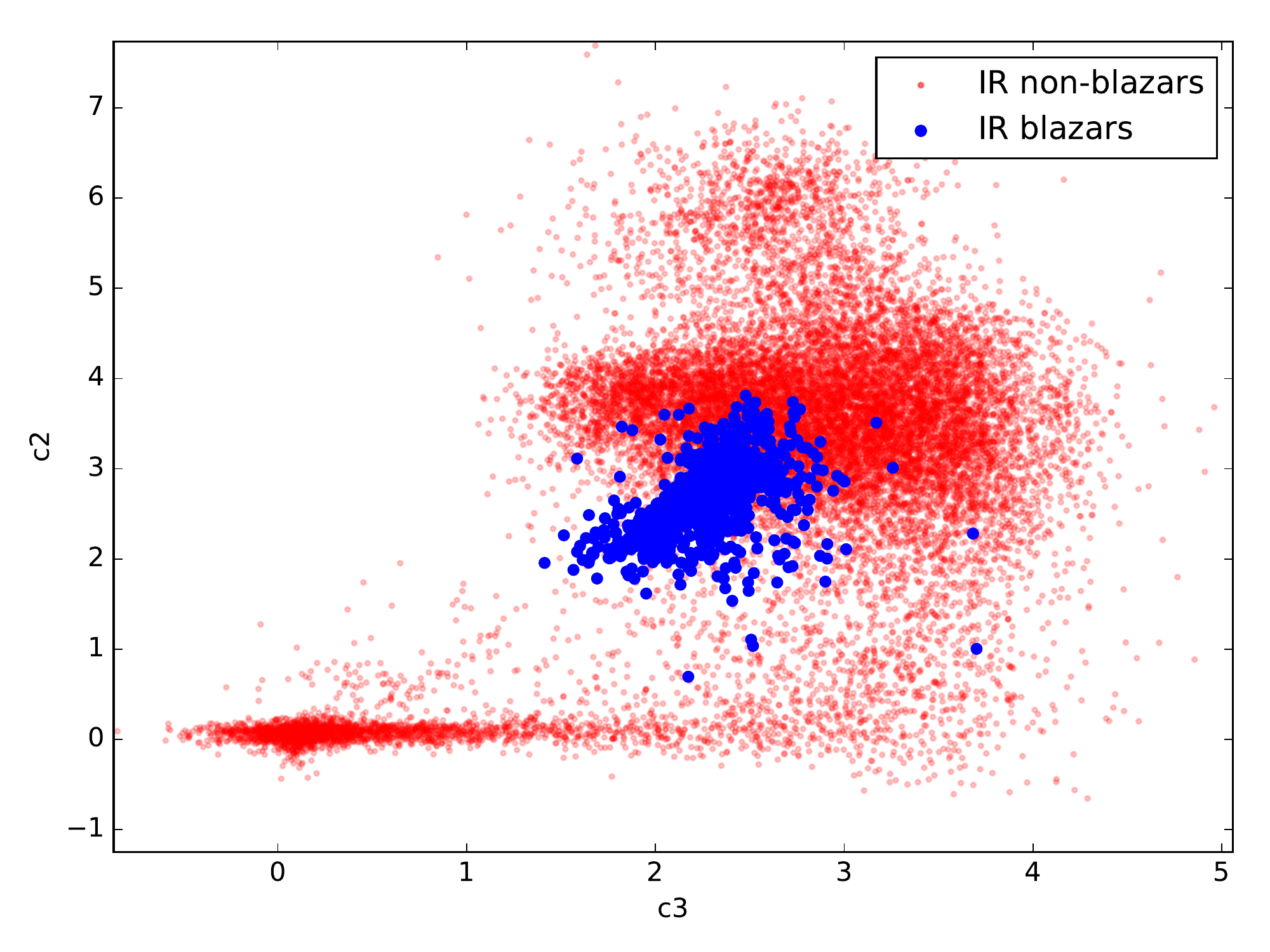}
  }
  \subfloat[]{
    \label{fig:c1c3}
    \includegraphics[width=0.33\columnwidth]{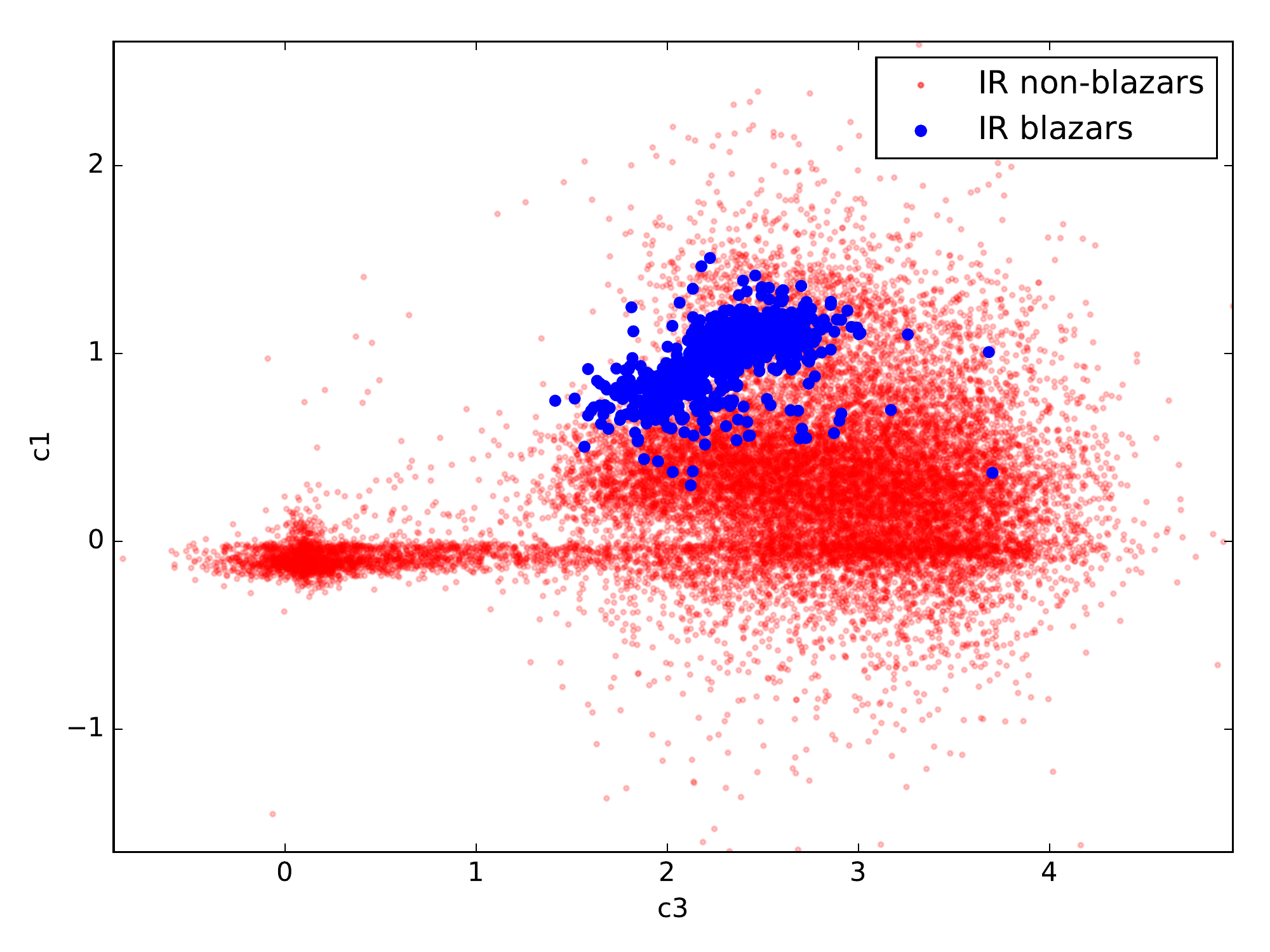}
  }\\ [-2ex]
  
  \caption{Scatter plots for (a) $c_1$ as a function of $c_2$,
    (b) $c_2$ as a function of $c_3$ and (c) $c_1$ as a function of $c_3$,
    for the infrared blazars (red) and the infrared non-blazars (blue).}
  \label{fig:colors}
\end{figure}

\begin{figure}[t]
  \centering
  \subfloat[]{
    \label{fig:rflux_ditribution}
    \includegraphics[width=0.48\columnwidth]{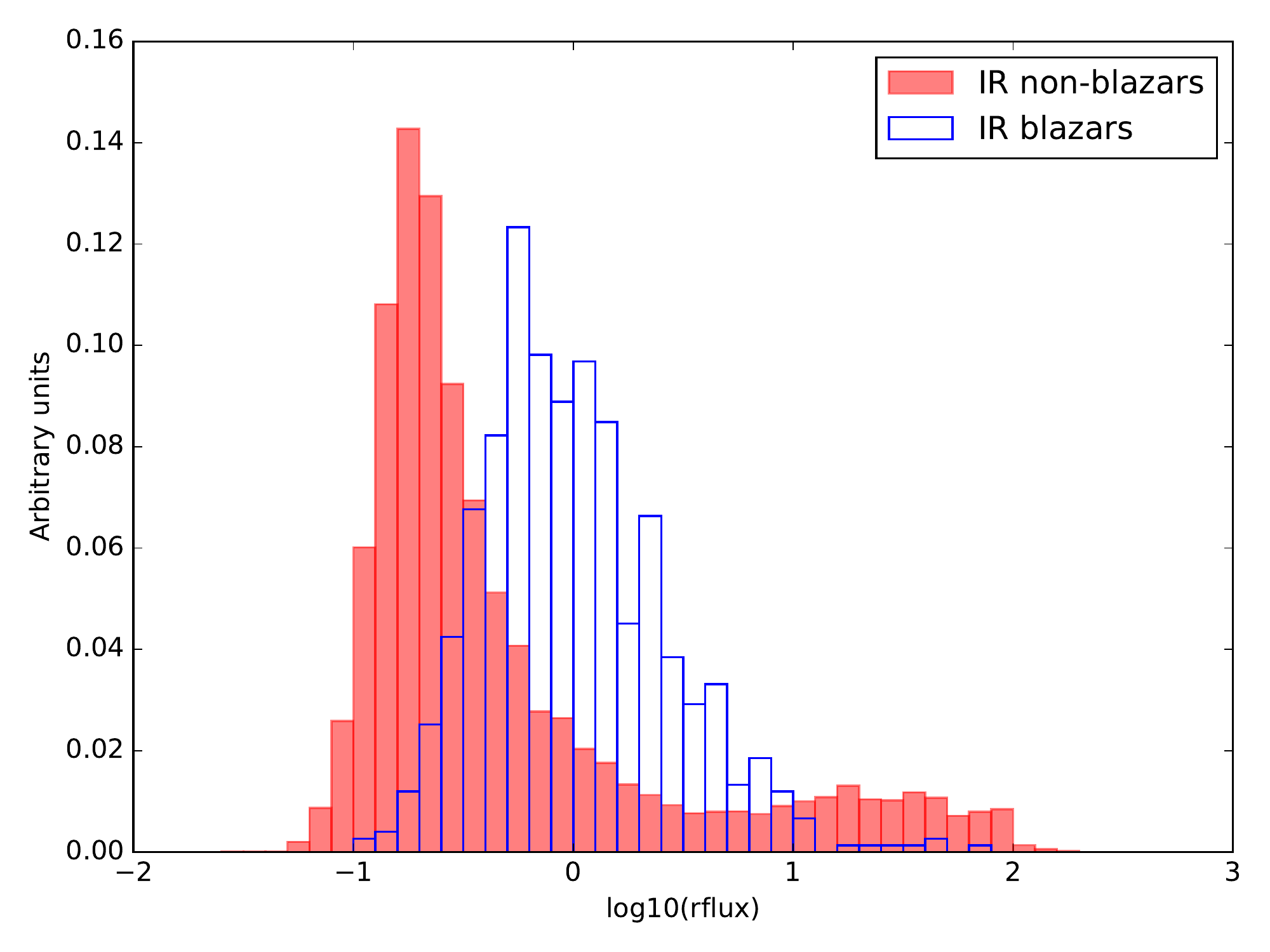}
  }
  \subfloat[]{
    \label{fig:rflux_redshift}
    \includegraphics[width=0.48\columnwidth]{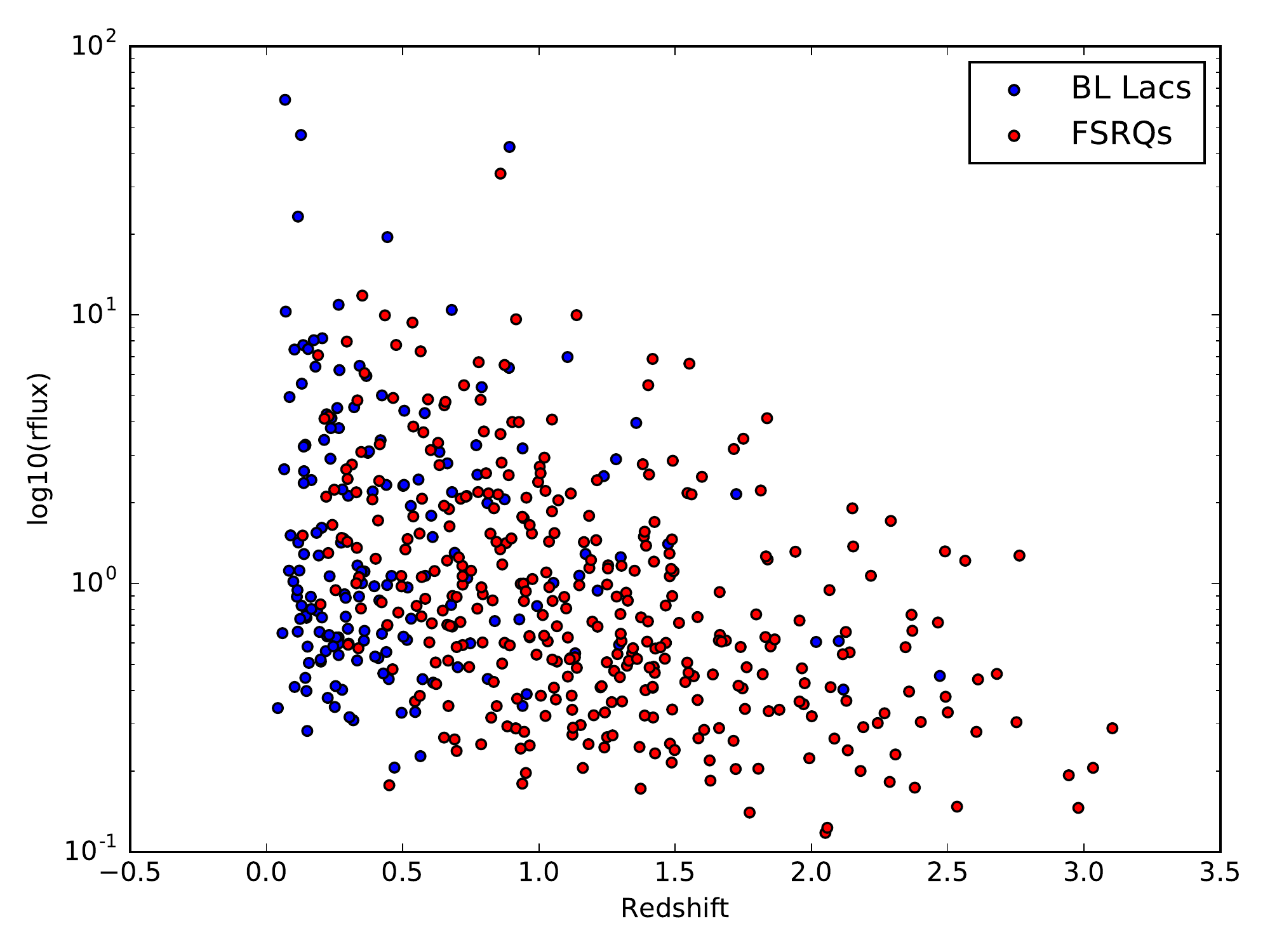}
  }\\ [-2ex]
  
  \caption{(a) Distribution of the $rflux$ parameter for the infrared
    blazars (red) and the infrared non-blazars (blue).
    (b) $rflux$ parameter as a function of the redshift for infrared blazars
    having a measured of redshift. BL Lacs and FSRQs are respectively shown
    in blue and in red.}
  \label{fig:rflux}
\end{figure}
To discriminate between the different classes of sources we used
the three colors $c_1 = W_1-W_2$, $c_2=W_2-W_3$ and $c_3=W_3-W_4$ as defined in
\citep{Massaro:2013ab}.
A sample of scatter plots is shown on Figure~\ref{fig:colors}.
In addition, we use a fourth parameter, called $rflux$, which quantify,
as a first order approximation, the blazars' exceptionally efficient
production mechanism of infrared photons compared to other classes of source.
This parameter is defined as the integrated flux estimated according to \cite{DAbrusco:2012aa}
divided by a value of reference corresponding to the average blazar integrated flux.
To study the Compton Dominance\footnote{Here we can not use CD as a discriminant parameter
  since we do not have an estimate of the $\gamma$-ray flux for the non-blazar infrared
  counterparts.} (CD), the ratio of luminosities between
the inverse-Compton and the synchrotron bumps, of a sample of blazars detected
by WISE and the Fermi/LAT, D'Abrusco et al. \cite{DAbrusco:2012aa} estimated
the integrated flux in the WISE energy range by summing the fluxes in the
four WISE filters.
Figure~\ref{fig:rflux_ditribution} shows the distribution of $rflux$.
The separation power between blazar and non-blazar infrared sources
is manifest but it is always tricky to use a flux as a discriminant parameter
as it is distance-dependent.
However, Figure~\ref{fig:rflux_redshift} shows the $rflux$ parameter
as a function of the redshift of 533 infrared blazars having an estimation of
redshift and we notice a small decrease of the flux which stays relatively
small compared to the bulk of the distribution.

\section{Classifier construction and association procedure}
\label{sec:train}
% BDT
% 70/30 %
% selection du split le + proche des perfs moyenne
% construction du modele avec la methode k-cv
% different point de fonctionnement.
% Plot ROC
% limitation du taux de contamination pour chaque association
% definition des classes pour lesquels le taux de contamination est
\begin{figure}[t]
  \centering
  \subfloat[]{
    \label{fig:roc}
    \includegraphics[width=0.48\columnwidth]{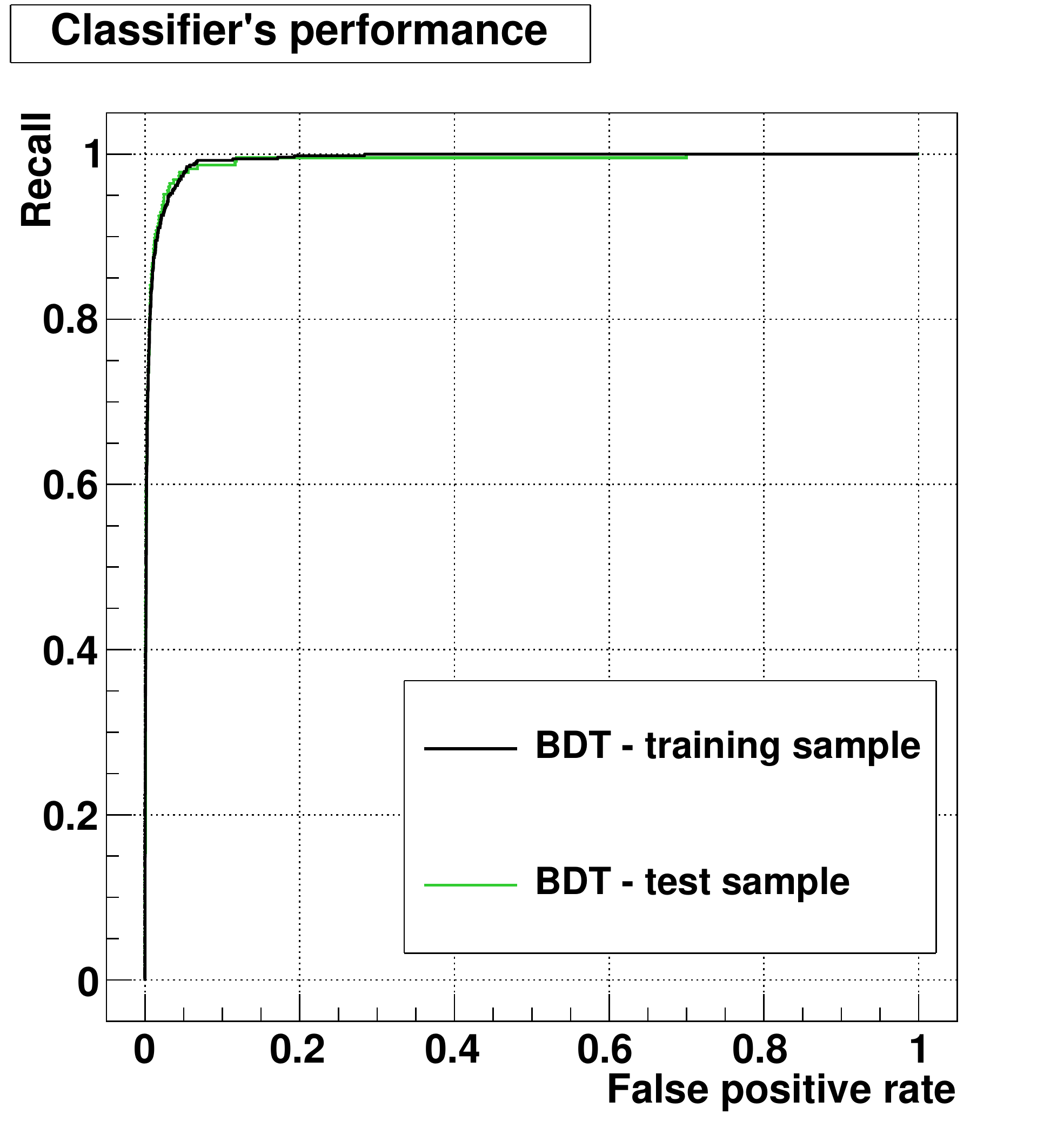}
  }
  \subfloat[]{
    \label{fig:nc}
    \includegraphics[width=0.48\columnwidth]{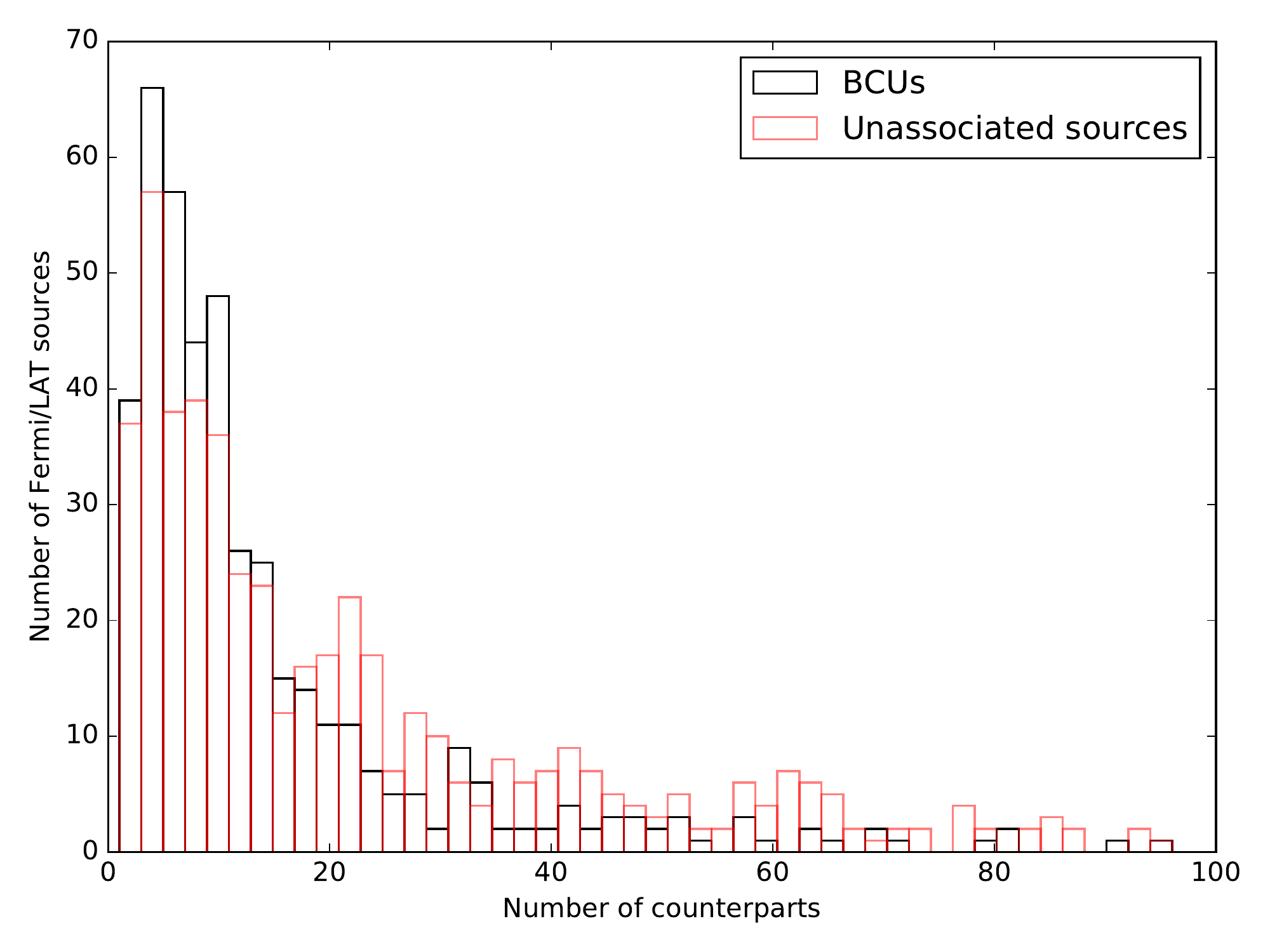}
  }\\ [-2ex]
  
  \caption{(a) The ROC curves of the boosted
    decision tree classifier estimated on the training sample using a ten-fold
    cross-validation method (black) and estimated on the test sample (green).
    (b) The number of infrared counterparts in the ``source'' region for
    the Fermi/LAT BCUs (black) and unassociated sources (red).
    The x-axis have been truncated above a number of 100 counterparts
    for readability.}
  \label{fig:perf}
\end{figure}
We considered several machine-learning algorithms to identify
potential infrared blazar-like counterparts of the Fermi/LAT sources.
We chose the boosted decision tree (BDT) algorithm for its capacity
to obtain good performance ``out of the box'', its robustness against over-training
and also for the smaller time needed to build a model compared to other methods
such as the neural-networks or the support vector machine based-methods.
We selected a single split of infrared blazars and non-blazars,
respectively \SI{70}{\percent} and \SI{30}{\percent} for the training and the
test samples.
The training/test split was obtained according to the proximity of its performance
compared to the average behaviour of the BDT method estimated on a large (100)
number of training/test phases with random splits
(see \cite{Lefaucheur:2017aa}).
To build a model from the training sample we used a ten-fold cross validation method.
The receiver operating characteristic (ROC) curves of the classifier
estimated on the training and on the test samples are shown on Figure~\ref{fig:roc}.
We determined the cutoff, called $\zeta^{\star}$, on the score distribution,
by requiring a true positive rate of \SI{90}{\percent}.
The performance metrics, the recall $\text{TP}$ and
the false positive rate $\text{FP}$,  estimated on the test sample
are respectively equal to \SI{91.19}{\percent} and \SI{1.66}{\percent}.
In the following, all infrared counterparts which score (called $\zeta$)
is greater or equal to $\zeta^{\star}$ will be considered as potential
blazar-like counterparts.

In order to control the number of expected false positives for each association, 
we defined classes of sources according to the expected contamination.
Each of the Fermi/LAT sources in the target samples has its own number of
infrared counterparts $N_{\text{c}}$, see Figure~\ref{fig:nc}, which we further
considered as dominated by non-blazar sources.
For each of the potential blazar-like counterparts one can
define an estimation of the expected number of false positives
$\eta=N_{\text{c}} \times  \text{FP}$, where the $\text{FP}$ value
is estimated with the score $\zeta$ of the source.
From this, we defined four classes of sources A, B, C and D which distribute the
infrared counterparts according to the expected number of false positives $\eta$:
\begin{itemize}
\item Class A, $\eta \leq 5/100$
\item Class B, $5/100 < \eta \leq 10/100$
\item Class C, $10/100 < \eta \leq 25/100$
\item Class D, $25/100 < \eta \leq 50/100$
\end{itemize}
In the following, the infrared candidates with an expected number of source
contamination greater than \SI{0.5}{} will not be considered for further analysis.

%% In order to obtain conservative estimates of the expected number of false
%% positives for each association in the error-box
%% of a Fermi/LAT source, we defined four regimes, associated to a false positive
%% rate, called $\text{FP}$, according to the score of an infrared source
%% We determined several cutoffs $\zeta^{\star}$ on the resulting score distribution,
%% called $\zeta$, by requiring true positive rates of \SI{90}{\percent},
%% \SI{80}{\percent}, \SI{70}{\percent} and \SI{50}{\percent}.
%% The performance on the test sample are summarised in the
%% Table~\ref{tab:perf} as long as the corresponding cutoff values.
%% \begin{table}
%%   \centering
%%   \begin{tabular}{|c|c|c|c|c|}
%%     \hline
%%     Regime & I & II & III & IV \\
%%     \hline
%%     TP$^{\text{cv}}$ (\SI{}{\percent})&
%%     50 &
%%     \SI{70}{} &
%%     \SI{80}{} &
%%     \SI{90}{} \\
%%     \hline
%%     $\text{TP}$ (\SI{}{\percent}) & 44.49 & 61.67 & 75.33 & 91.19 \\
%%     \hline
%%     $\text{FP}$ (\SI{}{\percent}) & 0.11 & 0.21 & 0.48 & 1.66 \\
%%     \hline
%%     $\zeta^{\star}$ & 0.2242 & 0.1362 & 0.0498 & -0.0822 \\
%%     \hline
%%   \end{tabular}
%%   \caption{Haha}
%%   \label{tab:perf}
%% \end{table}

\section{Application on the target samples}
\label{sec:appli}
%% For each of the two target samples, the unassociated sources and the BCUs,
%% we selected all the infrared sources in the Fermi/LAT error box, called the
%% ``source region'', defined as a circular region centered on the LAT
%% position with a radius of $\theta_{95}$, corresponding to the confidence
%% interval on the $\gamma$-ray source localisation at a level of \SI{95}{\percent}.
%% As a control, we also selected the infrared sources in an annular region
%% centered on the source region of internal and external radius of
%% $\theta_{95}$ and $\sqrt{2}\theta_{95}$, respectively, called the ``control''
%% region.

% BCUs
The application of the procedure to the 444 BCUs
of the 3FGL catalogue gives a total of 315 infrared blazar-like
counterparts for 265 Fermi/LAT sources ($\sim$1.2 counterparts
per $\gamma$-ray source).
The infrared sources are distributed among 197 Class A, 54 Class B,
39 Class C and 25 Class D corresponding to a number of expected
false positives less than 0.05, 0.10, 0.25 and 0.5, respectively.
%% For all the control regions of the 265 Fermi sources,
%% we found 63 counterparts for 53 $\gamma$-ray sources distributed
%% in 12 class A, 12 class B, 24 class C and 15 class D.
% Unids
The same procedure applied to the 531 unassociated sources
of the 3FGL catalogue gives a total of 188 infrared blazar-like
counterparts for 155 Fermi/LAT sources ($\sim$1.2 counterparts
per $\gamma$-ray source), distributed among 50 Class A, 35 Class B,
54 Class C and 49 Class D.
%% On the corresponding 155 control regions we obtained a total of 75 infrared
%% counterparts for 57 $\gamma$-ray sources.
%% Respectively 12, 12, 25 and 26 infrared counterparts
%% belongs to the class A, B, C and D.

\section{Discussion and conclusions}
\label{sec:end}
%% Accord avec le fond et sa maitrise
With our approach, building a classifier with a well-identified
sample of infrared blazars and a sample of infrared non-blazar sources,
we propose blazar counterparts for the Fermi/LAT sources and we can estimate
for each association an expected number of false positives.
%% we can estimate the number of false positives for each association, 
%% thanks to the performance metrics of the classifier.
Considering only the most promising associations, the source belonging to
class A or class B and correspond to a number of false positives less than
0.05 and 0.01, respectively, we find 251 potential blazar-like counterparts
for 235 BCUs.
The sum of the expected numbers of false positive is less than 7.
Furthermore, only 11 associations are found for these sources
in the control regions for which an equal or a better association exists
(10 for class A and one for class B).
For the Fermi/LAT unassociated sources, we find 85 blazar-like counterparts
of class A or B for 82 $\gamma$-ray sources with an expected number of false
positives less than 4.
In the corresponding control region, 9 counterparts have an equal or a
better association than the blazar-like counterparts in the signal region.

To assess the true nature of the selected infrared sources,
as candidates for an infrared counterpart of $\gamma$-ray blazar,
a multi-wavelength study is necessary.
In addition to provide astrometric coordinates to simplify the follow-up
at other wavelengths, it can help to prioritise the search for blazars.
For example, by crossing the list of blazar candidates for the unassociated
sources proposed in \citep{Lefaucheur:2017aa} and \citep{Saz-Parkinson:2016aa}
we found out that 130 infrared sources proposed in this work have a match,
including 60 infrared sources of class A or B.

In this work we focused on the high latitude Fermi/LAT sources from the
3FGL catalogue.
Dedicated classifiers should be used to tackle $\gamma$-ray sources of lower galactic
latitude, as the infrared source population differs from the high latitude ones.
Finally, the procedure could be applied to the Fermi/LAT 3FHL
catalogue \citep{The-Fermi-LAT-Collaboration:2017aa} in order to help the science
preparation with the upcoming of the Cherenkov Telescope Array.

\acknowledgments
This study used TMVA\footnote{\protect\href{http://tmva.sourceforge.net/}{http://tmva.sourceforge.net/}} \citep{Hoecker:2007aa}: an open-source toolkit
for multivariate data analysis.
STILTS\footnote{\protect\href{http://www.starlink.ac.uk/stilts/}{http://www.starlink.ac.uk/stilts/}} \citep{Taylor:2006aa} was used to manipulate tabular data, along with Astropy\footnote{\protect\href{http://www.astropy.org/}{http://www.astropy.org/}} \citep{Astropy-Collaboration:2013aa}, to fetch data, 
to cross match catalogues and to apply the corrections for the infrared extinction.
In addition, PyVO\footnote{\protect\href{https://pyvo.readthedocs.io/en/latest/}{https://pyvo.readthedocs.io/en/latest/}}
was used to fetch images from the WISE satellite.

%% Gros Travail multi-mwl a faire
%% Neanmoins, recouvrement avec d'autres etudes (D'abrusco et nous?)

%% Ouverture
% classifications dédiées pour différentes régions du ciel 
% contamination des AGN non-radio
% Choix du sample de blazar IR
% Etude MWL des candidats

% 

%\bibliographystyle{aa}
%\bibliographystyle{aipnum-cp}
\bibliographystyle{JHEP}
\bibliography{biblio}

\end{document}